\documentclass[%
 reprint,
 amsmath,amssymb,
 aps,
]{revtex4-2}

\usepackage{graphicx}% Include figure files
\usepackage{dcolumn}% Align table columns on decimal point
\usepackage{bm}% bold math
\begin{document}

\preprint{APS/123-QED}

\title{Blast waves in a paraxial fluid of light }% Force line breaks with \\

\author{Murad Abuzarli}
\author{Tom Bienaim\'e}
\author{Elisabeth Giacobino}
\author{Alberto Bramati}
\author{Quentin Glorieux}
 \email{quentin.glorieux@lkb.upmc.fr}
\affiliation{%
 ALaboratoire Kastler Brossel, Sorbonne Universit\'{e}, CNRS, ENS-Universit\'{e} PSL, Coll\`{e}ge de France - Paris, France
}%

\date{\today}% It is always \today, today,

\begin{abstract}
We study experimentally blast wave dynamics on a weakly interacting fluid of light.
%In this regime, the perturbation density which is only several times larger than the background.
The fluid density and velocity are measured in 1D and 2D geometries.
Using a state equation arising from the analogy between  optical propagation in the paraxial approximation and the hydrodynamic Euler's equation, we access the fluid hydrostatic and dynamic pressure.
In the 2D configuration, we observe a negative differential hydrostatic pressure after the fast expansion of a localized over-density, which is a typical signature of a blast wave for compressible gases.
Our experimental results are compared to the Friedlander waveform hydrodynamical model\cite{friedlander1946diffraction}.
Velocity measurements are presented in 1D and 2D configurations and compared to the local speed of sound, to identify supersonic region of the fluid.
Our findings show an unprecedented control over hydrodynamic quantities in a paraxial fluid of light.
\end{abstract}

\maketitle

%\tableofcontents

\section*{Introduction}
In classical hydrodynamics, a blast wave is characterized by an increased pressure and flow resulting from the rapid release of energy from a concentrated source \cite{Dewey2016}.
The particular characteristics of a blast wave is that it is followed by a wind of negative pressure, which induces an attractive force back towards the origin of the shock.
Typical blast waves occur after the detonation of trinitrotoluene \cite{dewey1964air,reed1977atmospheric}, nuclear fission \cite{taylor1950formation},  break of a pressurized container \cite{xu2017dispersive} or heating caused by a focused pulsed laser \cite{li2004characteristics}.
The sudden release of energy causes a rapid expansion, which in a three dimensional space is analogous to a spherical piston \cite{PhysRevLett.100.084504} and produces a compression wave in the ambient gas. 
For a fast enough piston, the compression wave develops into a shock wave which is characterized by the rapid increase of all the physical properties of the gas, namely, the hydrostatic pressure, density and particle velocity \cite{dewey2018friedlander}.
In 1946, Friedlander predicted that immediately after the shock front the physical properties at a given position in space decay exponentially \cite{friedlander1946diffraction,dewey2010shape}.
In this model, for 3-dimensional and 2-dimensional spaces the hydrostatic pressure and the density are expected to fall below the values of the ambient atmosphere leading to a blast wind \cite{Dewey2016}.

Shock waves have been studied in several contexts in physics, including acoustics, plasma physics, ultra-cold atomic gases \cite{PhysRevA.74.023623,PhysRevA.80.043606,PhysRevLett.101.170404} and non-linear optics \cite{wan2007dispersive,vocke2015experimental,conti2009observation,ghofraniha2012shock,xu2017dispersive}.
In optics, the hydrodynamics interpretation relies on the Madelung transforms which identify the light intensity to the fluid density and the phase gradient to the fluid velocity\cite{IC2014superfluid}.
Recently several works have studied analytically shock wave formation in one and two dimensions \cite{MIsoard_WB19,Tom2021controlled}.
Optical systems allow for repeatable experiments and precise control of the experimental parameters.
For example dispersive superfluid-like shock waves have been observed \cite{wan2007dispersive}, as well as generation of solitons \cite{conti2009observation}, shocks in non-local media \cite{ghofraniha2007shocks,vocke2015experimental}, shocks in disordered media \cite{ghofraniha2012shock}, analogue dam break \cite{xu2017dispersive} and Riemann waves \cite{wetzel2016experimental}.
However, an experimental study of blast waves has not been done in atomic gases nor in non-linear optics systems.
In this work, we demonstrate the generation of a blast wave in a fluid of light.
Interestingly, the prediction of a blast wind with negative pressure and density holds in two dimensional space but not in 1 dimension \cite{sadot2018small}. 
Optical analogue systems allow for an experimental validation of this prediction.

In this letter, we study the formation of blast waves in a paraxial fluid of light.
We measure the time evolution of analogue physical properties such as the hydrostatic pressure, the density, the particle velocity and the dynamic pressure at a fixed point for 1 and 2-dimensional systems.
We report the observation of a negative hydrostatic differential pressure after a shock wave in 2-dimensional system and we show that the Friedlander waveform describes quantitatively our experimental results for all physical parameters.
This paper is organized as follows.
We first introduce the analogy between the propagation equation of a laser beam through non-linear medium (a warm atomic vapor) and the hydrodynamics equation and derive the relevant analogue physical properties.
In the second section of this work, we describe our experimental setup and present our results on the density and hydrostatic pressure measurements.
We highlight the striking differences between 1 and 2-dimensional systems.
Finally, we study the time evolution of the velocity and dynamic pressure.

\section*{Theoretical Model}

% - Paraxial fluid of light/ effective time / adimensional/introduce velocity  : Murad
% - Tau vs z/znl : evolution = changing power : Murad
% (pas de SW ici)
% -SW and blast : QG

We describe the propagation of a linearly polarized monochromatic beam in a local Kerr medium. We separate the electric field's fast oscillating carrier from the slowly varying (with respect to the laser wavelength) envelope : $E=\mathcal{E}(\mathbf{r},z)$e$^{i(kz-k_0c t)}+$ complex conjugate. Under the paraxial approximation, the propagation equation for the envelope $\mathcal{E}$ is the Non-Linear Schrödinger Equation (NLSE) \cite{IC2014superfluid}:
\begin{equation}
    i\frac{\partial\mathcal{E} }{\partial z}=
    \left(-\frac{1}{2k}\nabla^2_{\perp}+g{\mid}\mathcal{E}{\mid}^2-\frac{i\alpha}{2}
    \right)\mathcal{E},
\end{equation}
where $k$ is the laser wavevector in the medium, $\alpha$ is the extinction coefficient accounting for losses due to absorption, and the $g$ parameter is linked to the  intensity dependent refractive index variation $\Delta n $  via: $g{\mid}\mathcal{E}{\mid}^2=-k_0\Delta n$ (with $k_0$ the laser wavevector in vacuum).

The NLSE is analogous to a 2D Gross-Pitaevskii equation describing the dynamics of a quantum fluid in the mean-field approximation.
This analogy is possible by  mapping the envelope $\mathcal{E}$ to the quantum fluid many-body wavefunction and the axial coordinate $z$ to an effective evolution time.
The non-linear refractive index variation plays then the role of a repulsive photon-photon interaction, since all measurements in this work are done in the self-defocusing regime i.e. $\Delta n<0$ and therefore $g>0$.
Diffraction acts as kinetic energy with the effective mass emerging from the paraxial approximation and given by the laser wavevector $k=8.10^6$ m$^{-1}$.
Using the Madelung transformation: $\mathcal{E}=\sqrt{\rho}\textrm{e}^{i\phi}$, $\mathbf{v}=\frac{c}{k}\nabla_{\perp}\phi$ one can derive from the NLSE hydrodynamic equations \cite{MIsoard_WB19, wan2007dispersive}, linking the fluid's density $\rho$ with its velocity $\mathbf{v}$:
\begin{eqnarray}
    \frac{\partial\rho}{\partial z}+
    \nabla_{\perp}.\left(\rho \frac{\mathbf{v}}{c}\right)=-\alpha\rho
    \label{Continuity}\\
    \frac{\partial\mathbf{v}}{\partial z}+
    \frac{1}{2c}\nabla_{\perp}\mathbf{v}^2=
    -\nabla_{\perp}\left(\frac{cg\rho}{k}-
    \frac{c}{2k^2\sqrt{\rho}}\nabla^2_{\perp}\sqrt{\rho}
    \right). \label{Euler}
\end{eqnarray}
Eq. (\ref{Continuity}) is the continuity equation with a loss term accounting for photon absorption.
Eq. (\ref{Euler}) is similar to the Euler equation without viscosity, in which the driving force stems from interaction and the so-called quantum pressure term due to diffraction. 
Establishing the formal analogy requires, however, defining an analogue pressure $P$ to be able to re-express the right-hand side of Eq. (\ref{Euler}) as: $-1/\rho\cdot\nabla_{\perp}P$.
This is possible for the first term stemming from interactions.
Using the identity: $ -\nabla_{\perp}\rho=-1/(2\rho)\nabla_{\perp}\rho^2$ one can define the so called bulk hydrostatic pressure $P$ as:
\begin{eqnarray}
    P=\frac{c^2}{2}\frac{\rho^2 g}{k}=\frac{1}{2}\rho c_s^2,
    \label{PresDef}
\end{eqnarray}
where the last equality comes from $c_s^2=c^2\cdot g\rho/k$.
Eq. (\ref{PresDef}) is the state equation linking the fluid hydrostatic pressure $P$ to its density if one neglects the quantum pressure term. 
It is the consequence of the mean-field formulation of the interaction.
It also implies that the fluid of light is compressible with the compressibility equal to: $k/(c^2\rho^2g)$.
One then gets the analogue Euler equation: 
\begin{eqnarray}
    \frac{\partial\mathbf{v}}{\partial (z/c)}+
    \frac{1}{2}\nabla_{\perp}\mathbf{v}^2=
    -\frac{1}{\rho}\nabla_{\perp}P, \label{Euler1}
\end{eqnarray}
with a pressure $P$ of dimension [\textit{density}]$\times$[\textit{speed}]$^2$. 
As already mentioned, the fluid dynamics can be studied by accessing its state at different $z$ positions, however this is not recommend practically since imaging inside a non-linear medium is highly challenging task. 
Alternatively, one can instead re-scale the effective time by incorporating fluid interaction \cite{Pavloff_Cargese_2019,Tom2021controlled}.
Fluid interaction can then be varied experimentally and the fluid dynamics can be studied  while imaging only the state at the medium output plane.
Re-scaling the time is based on defining following quantities: 
\begin{eqnarray}
    z_{NL}=\frac{1}{g\rho(0,L) }, \ \ \textrm{non-linear axial length} \label{z_NL}\\
    \xi= \sqrt{\frac{z_{NL}}{k},} \ \ \textrm{transverse healing length} \label{xi} \\
    c_s=\frac{c}{k\xi}, \ \ \textrm{speed of sound} \label{c_s} \\
    \psi=\frac{\mathcal{E}}{\sqrt{\rho(0,L)}}, 
\end{eqnarray}
and substituting the time and space variables as:
$\tau=~z/z_{NL}$, $\tilde{\mathbf{r}}=\mathbf{r}/\xi$,
$\tilde\nabla_{\perp} = \xi\nabla_{\perp}$.
$L$ is the non-linear medium length.
The propagation equation then reads:
\begin{equation}
    i\frac{\partial\psi }{\partial \tau}=
    \left(-\frac{1}{2}\tilde\nabla^2_{\perp}+{\mid}\psi{\mid}^2
    \right)\psi.
    \label{GPE_Adim}
\end{equation}
One can note that dynamics of $\psi$ is not anymore dissipative, due to the normalization with respect to the exponentially decaying density: $\rho(0,L)=\rho(0,0)\textrm{exp}(-\alpha L)$, measured at at the medium exit plane.
This formulation is necessary to describe accurately the experimental results of this work probing temporal dynamics of a fluid of light by varying the strength of the optical non-linearity and not the imaged $z$ plane.
The effective time $\tau = |\Delta n(\mathbf{r}_\perp=0,L)| k_0L$ equals to the maximal accumulated non-linear phase. 
Rewriting the Madelung transformation with the new variables, we obtain:
\begin{equation}
    \psi=\sqrt{\tilde\rho}\textrm{e}^{i\phi}=\sqrt{\frac{\rho}{\rho(0,L)}}\textrm{e}^{i\phi}, \ \ \ \tilde{\mathbf{v}}=\frac{\mathbf{v}}{c_s}=\tilde\nabla_{\perp}\phi. \label{FlVelo}
\end{equation}
One gets dimensionless Euler and the continuity equations:
\begin{eqnarray}
    \frac{\partial\tilde{\rho}}{\partial \tau}+
    \tilde{\nabla_{\perp}}.\left(\tilde{\rho}\tilde{\mathbf{v}}\right)=0\\
    \label{ContinuityAdim}
    \frac{\partial\tilde{\mathbf{v}}}{\partial \tau}+
    \frac{1}{2}\tilde{\nabla_{\perp}}\tilde{\mathbf{v}}^2=-
    \tilde{\nabla_{\perp}}\left(
    \tilde{\rho}-\frac{1}{2\sqrt{\tilde{\rho}}}
    \tilde{\nabla^2_{\perp}}\sqrt{\tilde{\rho}}
    \right),
    \label{EulerAdim}
\end{eqnarray}
where the link between Eq. (\ref{EulerAdim}) and the Euler equation is made by neglecting the quantum pressure and defining the the dimensionless hydrostatic pressure as:
\begin{equation}
    \tilde {P}=\frac{1}{2}\tilde {\rho}^2 \label{PresAdim}.
\end{equation}
Finally, the dynamic pressure is defined as a vector quantity by:
\begin{equation}
    \tilde {P_d}=\frac{1}{2}\tilde  {\rho}  \tilde {\mathbf{v}} |\tilde {\mathbf{v}}|\label{PresDyna},
\end{equation}
The dynamic pressure is the fluid kinetic energy flux and accounts for the amount of pressure due to fluid motion.
The impact force on an obstacle hit by a shockwave is proportional to its dynamic pressure.
Expressed in dimensionless units, the dynamic pressure gives the strength of the convection term normalized by the pressure due to the interactions in the Eq. (\ref{EulerAdim}).
It can be computed directly from the density and velocity measurements.

\section*{Shock waves and blast wind}
In this work, we study the dynamics of a fluid of light disturbed by a localized Gaussian over-density $\delta\rho(\mathbf{r},0)=~\rho_1~\textrm{exp}\left(-2\mathbf{r}^2/\omega_1^2\right)$.
$\rho_1$ is of the same magnitude as the background fluid density $\rho_0$ and $\omega_1$ quantifies the perturbation width.
We can write $\rho(\mathbf{r},L)= \rho_0+\delta\rho(\mathbf{r},L).$
Normalizing the total density by its maximal undisturbed value one gets: $\tilde\rho(\mathbf{r},\tau)=\rho(\mathbf{r},L)/ \rho_0(0,L)$.
Extending this definition to $\rho_0$ and $\rho_1$, we obtain $\tilde\rho_0$ bound between 0 and 1 and having a Gaussian shape, and $\tilde\rho_1$ expressing the perturbation strength with respect to the fluid background density. 
To take into account the Gaussian profile of the density $\rho_0$, we define the over-pressure from the pressure difference between the case with and without perturbation:
\begin{equation}
    \delta\tilde P(\mathbf{r},\tau)= \tilde P(\mathbf{r},\tau) - \tilde P_0(\mathbf{r},\tau). \label{OverPressure}
\end{equation}
To evaluate the differential pressure $\Delta\tilde P(\tau)$, showing the instantaneous difference in pressure between the perturbation center and the external undisturbed area, we define:
\begin{equation}
    \Delta\tilde P(\tau)= \tilde P(0,\tau) - P_0(r_{ext},\tau). \label{DiffPressure}
\end{equation}
The differential pressure $\Delta\tilde P(\tau)$ is the most important quantity we study in this work and we expect major differences in the non-linear perturbation dynamics between the 1D and the 2D geometries. 
Finally, the fluid velocity can be measured experimentally.
It requires a measurement of the beam wavefront which is realized using off-axis interferometry. 
Calculating numerically the gradient of the phase, we obtain the background fluid velocity $\mathbf{v}_0$ and the perturbation velocity  $\mathbf{v}_1$ by analyzing the images without and with the perturbation, respectively.

Several studies have been performed in both $\rho_1\ll~\rho_0 $ and $\rho_1\gg\rho_0$ regimes, observing the Bogoliubov dispersion of the linearized waves created by the perturbation \cite{QF_Bogo18,fontaine2020interferences,piekarski2020short}, and the shock waves \cite{wan2007dispersive,Tom2021controlled}, respectively. 
In this work we investigate the case $\rho_1\sim\rho_0$ by analyzing the fluid density, velocity and pressure both in the 1D and 2D geometries. 
The NLSE is known to give rise to sound-like dispersion to the low amplitude waves, governed by the Bogoliubov theory. 
Here, a perturbation of the same order (or larger) than the background results in the sound velocity variation following the local density inside the perturbation. 
This is the prerequisite for observing shock waves, a special type of waves changing their shape during propagation towards a steepening profile.  
In hydrodynamics, shock waves are usually reported as a time evolution measurement of a physical quantity (pressure, density...) at a fixed point in space.
After the passage of a the shock wave front, a blast wind (a negative differential pressure) should be observed in 2 and 3 dimensional space.
A direct physical consequence of this wind in classical hydrodynamics is observed for example after an explosion inside an edifice: the presence of glass pieces within the building is the signature of the blast wind .
In the next section we report the time evolution as well as the time snapshots (spatial map of a physical quantity at fixed time) typically not accessible in classical hydrodynamics experiments.

\section*{Experimental setup}
In our experiment, we investigate the propagation of a near-resonance laser beam through a warm rubidium vapor cell, which induces effective photon-photon interactions \cite{agha2011time}.
Two configurations are studied: the 2D geometry with a radially symmetric dynamics and the 1D geometry with a background much larger along $x$ than along $y$ which allows for a 1D description of non-linear wave dynamics \cite{Tom2021controlled}.
A tapered amplified diode laser is split into a background, a reference and a perturbation beams (see supplementary materials for details). 
The background beam is enlarged with a telescope up to a waist of 2.5$\pm$0.5~mm along $x$ and 0.8$\pm$0.1~mm along $y$ in the 1D geometry, and 1.8$\pm$0.3~mm along the radial coordinate in the 2D configuration.
The reference beam (for interferometric phase measurement) is matched to the same dimensions.
The perturbation beam is focused to get the waist of 0.12$\pm$0.03~mm in the middle of the cell (the corresponding Rayleigh range is 55~mm). 
The  background and perturbation are recombined with a 90R:10T beam splitter such that 90 \% of the background beam power is reflected towards the cell. 
The second arm of the BS is sent through a 200 $\mu m$ diameter pinhole into a photodiode to stabilize the interferometer.
The control is realized by locking on local minimum acting on a piezoelectric mirror mount with a RedPitaya hardware run by the PyRPL software \cite{PyRPLref}. 
Cell temperature is 149(2)°~C  leading to an atomic density of 8.3$\pm$0.8$\times10^{13}$ cm$^{-3}$.
The cell output is imaged with a $\times$4.2 magnifying 4-f setup onto a camera.
Sets of 4 images (background only, background with reference, background with perturbation and finally background with both perturbation and reference) at different input powers $\mathcal{P}$ ranging from 50 to 600~mW and different laser detunings $\Delta$ from the $^{85}$Rb D2 line $F=3\rightarrow F'$ transition are taken (see supplementary materials for details).
The reference beam is superimposed with other beams with an angle of 30 milli-radians, giving rise to interferogramms with vertical fringes of average periodicity of 25$\pm$1~$\mu$m.

\begin{figure}[]
\includegraphics[width=0.89\columnwidth]{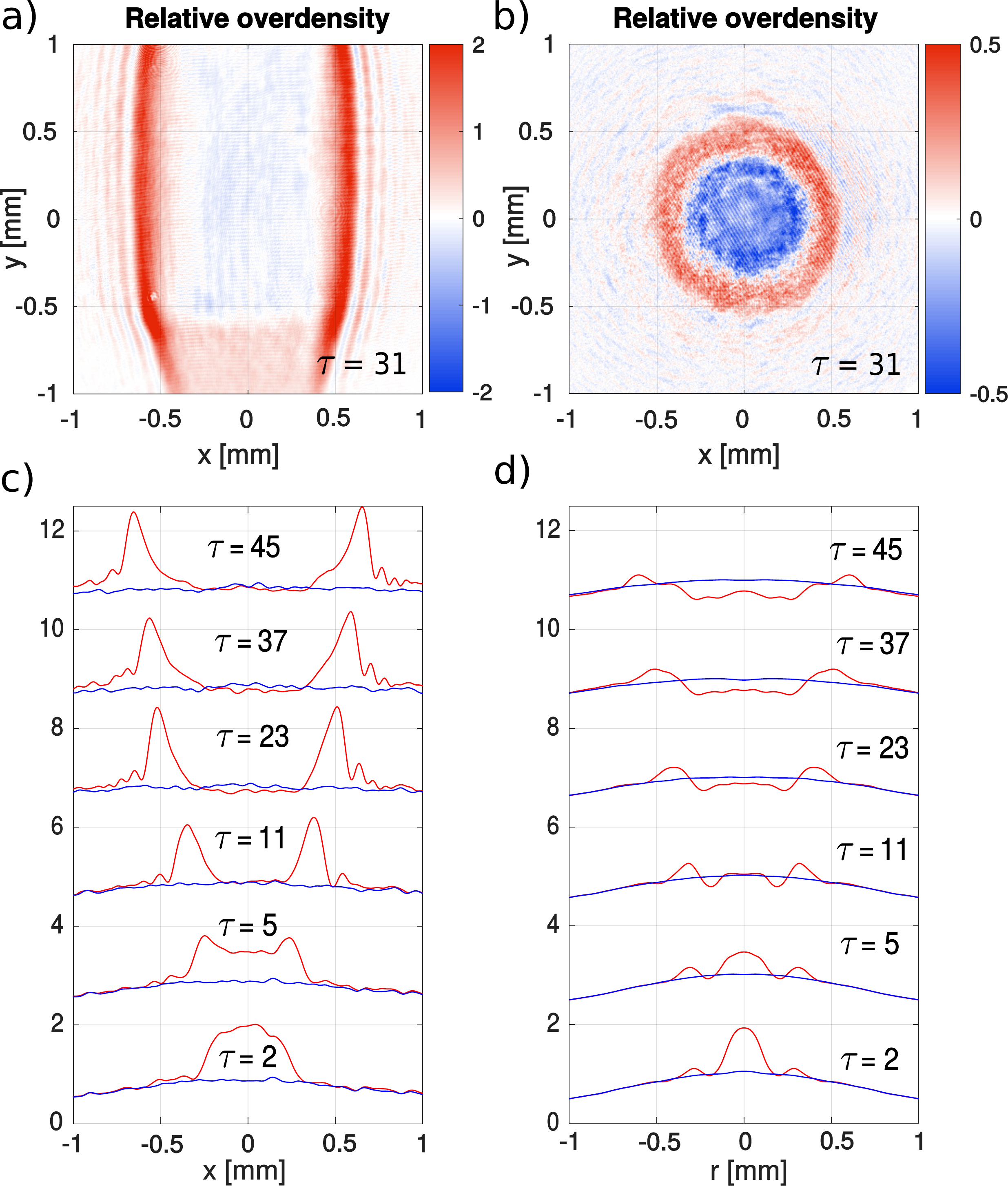}
\caption{Density data: The left column corresponds to the 1D configuration and the right column to the 2D case.
a) and b) are over-density maps at time $\tau=31$, obtained by subtracting the images with no perturbation from the ones with perturbation in the 1D and 2D geometry, respectively. 
c), d) are density profiles without (blue) and with (red) perturbation in the 1D and 2D geometry, respectively. 
The profiles are shifted vertically (spacing of 2) for better visibility.}
\label{fig1_density}
\end{figure}

\section*{Density}
The density is an important physical parameter needed to compute the static and hydrodynamic pressure.
It is directly given by the intensity measurement.
In figure~\ref{fig1_density} a) and b), we present the experimental maps of the over-density $\delta\tilde\rho$ at time $\tau=31$, after subtracting the background fluid, in the 1D and 2D geometries respectively. 
By changing the laser intensity and detuning, we can modify the effective time $\tau$.
The associated time $\tau$ is calculated from the nonlinear index $\Delta n$ via the off-axis interferometric measurement for each experimental configuration ($\mathcal{P}, \Delta$) (see supplementary materials for details).
Fig.~\ref{fig1_density} a) and b) show the spatio-temporal over-density diagrams. 
We present the corresponding density profiles at different times in figure~\ref{fig1_density} c) and d).
The 1D density data are averaged over the $y$ direction for ${\mid}y{\mid}<0.1$~mm and the 2D images are radially averaged to get the background fluid density (blue curves) and the total fluid density including the perturbation (red curves). 
These results show two important effects. 
In the 1D geometry, a clear steepening of the perturbation front and the development of dispersive shock waves can be seen as an oscillating pattern developing in beyond the shock front with effective time $\tau$.
In the 2D geometry, interestingly, the steepening of the shock front is less pronounced. 
Moreover, a density much lower than the background density is observed in the center of the 2D profiles for long time $\tau>20$, which is not the case in 1D.
This negative differential density has a direct consequence on the differential pressure calculated using Eq. (\ref{DiffPressure}).

\section*{Static pressure}

\begin{figure}[]
\includegraphics[width=0.89\columnwidth]{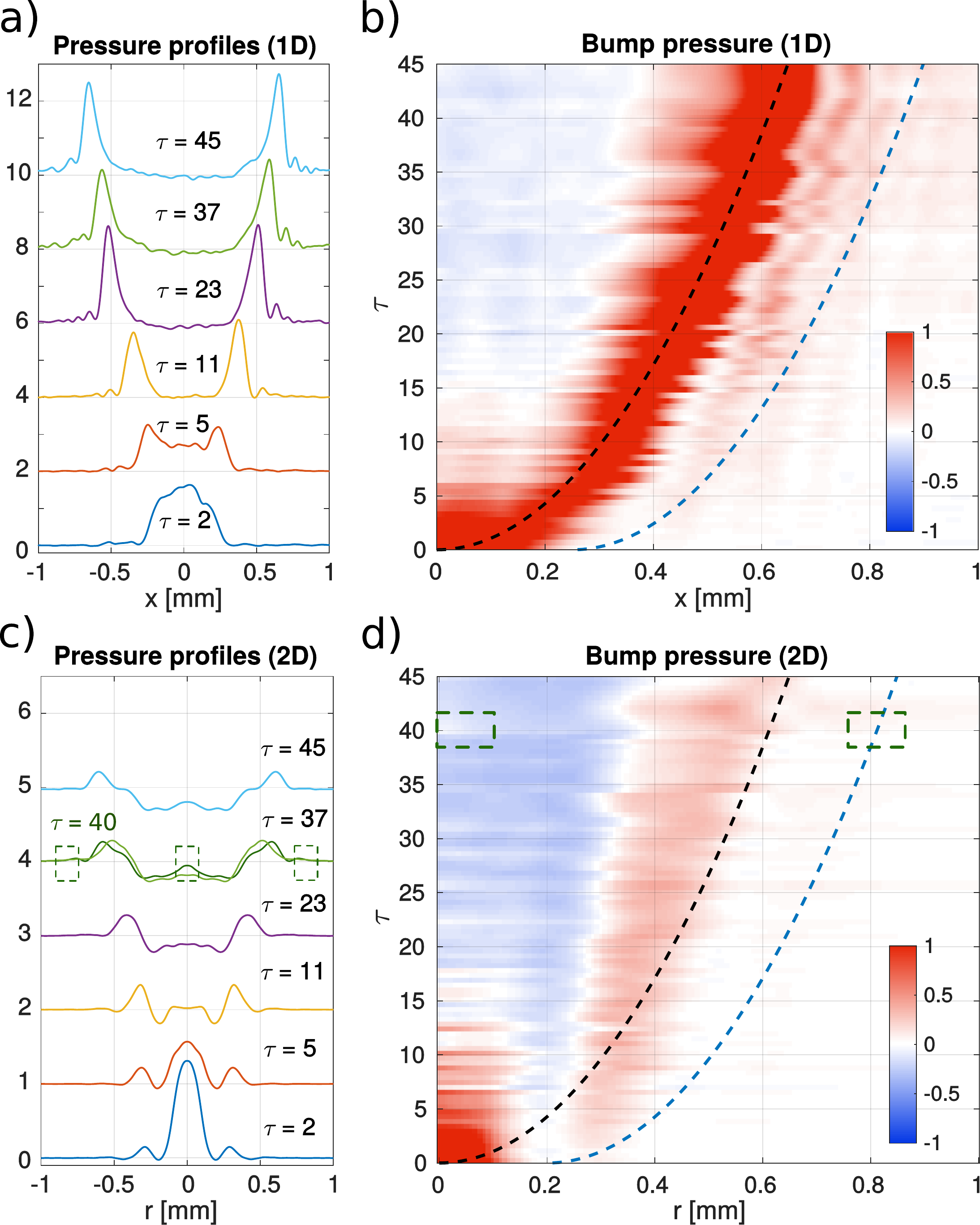}
\caption{Pressure analysis: a),c): 1D \& 2D over-pressure profiles evaluated at different effective times $\tau$. Each following profile shifted vertically by 2 for better visibility. b),d) show the 1D and 2D spatio-temporal diagrams of the over-pressure evolution, respectively. The dotted black lines show the trajectory of expansion at the speed of sound according to the parabolic equation with the prefactor given by $k/L=107$~mm$^{-1}$ in both geometries. The blue dotted lines show the same trajectories shifted horizontally by 250~$\mu m$ and 200~$\mu m$ in 1D and 2D cases, respectively. It corresponds to external undisturbed area used for the measurement of the differential pressure.
Dashed green rectangles around $\tau=40$ show the presence of a second shock due to an increasing differential pressure. }
\label{fig2_pressure}
\end{figure}

\begin{figure}
\includegraphics[width=0.8\columnwidth]{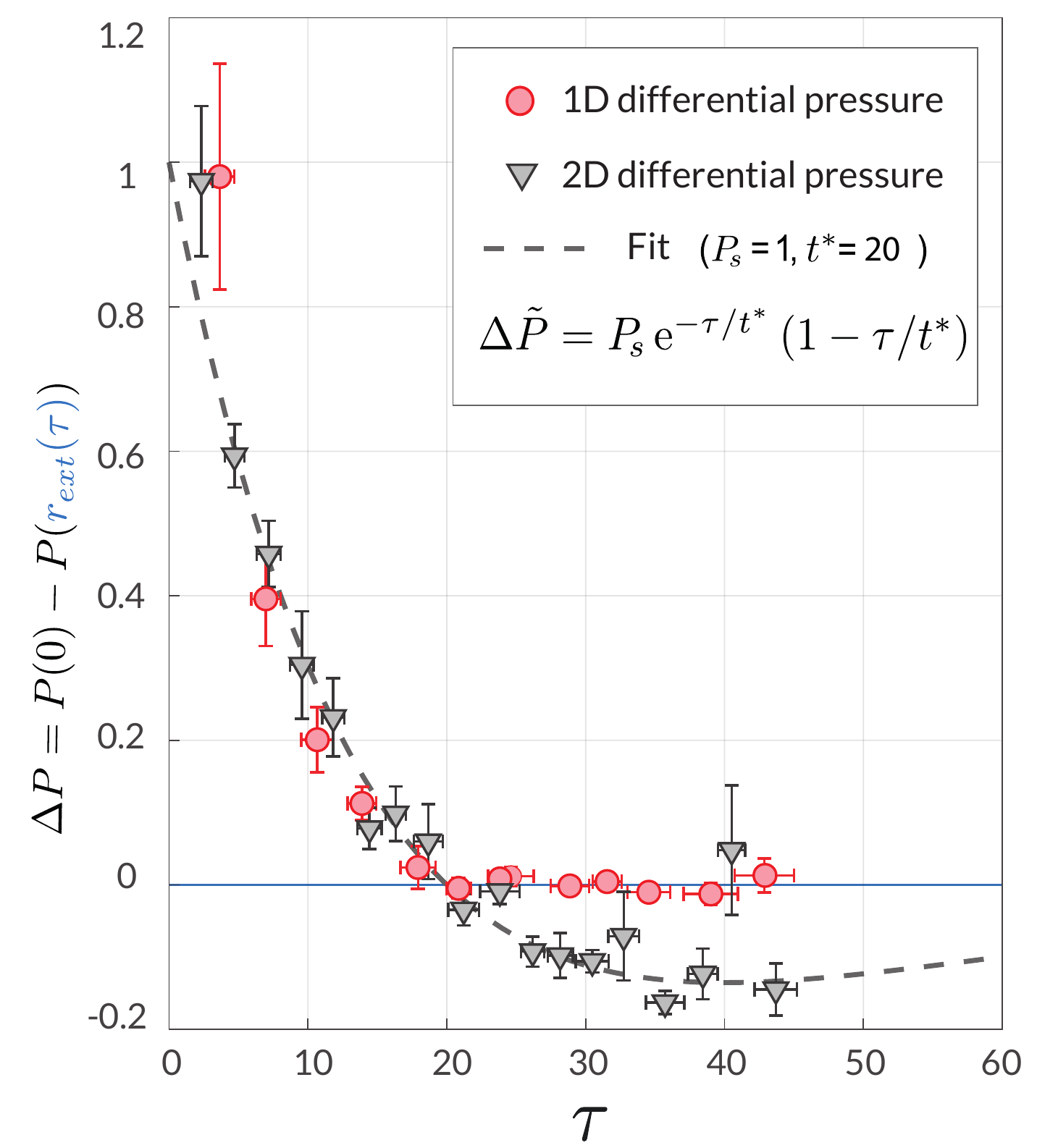}
\caption{Differential pressure calculated from Eq. (\ref{DiffPressure}) for the 1D (circular dots) and the 2D cases (square dots). The uncertainty bars correspond to the statistical analysis of multiple images. The pressure is normalized as described in the main text. Blue line is the ambient pressure outside of the shock. Black dashed line is the Friedlander model for a blast wave described in Eq. (\ref{friedland}) with $P_s=1$ and $t^*=20$. }
\label{fig3_fit}
\end{figure}

%discuss fig 2, and how to make fig 3 (differential pressure 95r2 etc) :MA
%Discuss Fit and 1D vs 2D : QG
To isolate the effect of the perturbation on the static pressure, we compute the over-pressure from images of the background with and without the bump taken at same effective times $\tau(\mathcal{P},\Delta)$, using Eq. (\ref{OverPressure}) and (\ref{PresAdim}).
The over-pressure as a function of time $\tau$ is shown in Fig.~(\ref{fig2_pressure}) b) and d) and profiles averaged along $y$ in the 1D case and radially in the 2D case are presented in Fig.~(\ref{fig2_pressure}) a) and c) for various times.

The trajectory of a density pulse spreading with no dispersion at the speed of sound can be expressed as follows: $r =c_s(\tau)\times (L/c)$.
The coefficient can be calculated using the time dependence of the sound velocity: $c_s=c\sqrt{\tau/(kL)}$ obtained from  Eqs. (\ref{z_NL}) and (\ref{c_s}). 
It directly leads to $\tau = kr^2/L$ and knowing that: $L=75$~mm and $k=8\times10^3$~mm$^{-1}$, one gets: $\tau=107\times r^2$.
The coefficient does not depend on the dimensionality of the system.

In the pressure maps (Fig.~(\ref{fig2_pressure}) b) and d)), we have added a black dashed line following this trend: $\tau=107\times x^2$ (1D) and $\tau=107\times r^2$ (2D).
As expected, this trajectory follows closely the shock front in the 1D geometry.
The differential pressure is defined as the pressure difference between inside and outside of the shock as expressed in Eq.~(\ref{DiffPressure}).
The undisturbed pressure as function of time is evaluated along the same trend line $\tau=107\times (r_{ext}-r_0)^2$, translated $r_0=250~\mu$m in 1D and $r_0=200~\mu$m in 2D, which corresponds to $\sim 1.5$ times the perturbation beam waist (blue dashed line).
In 2D, the shock front expansion is slower than the calculated trajectory, as described in \cite{Tom2021controlled}, and the blue dashed line can therefore still used to define the undisturbed pressure.

The temporal evolution of the differential static pressure (at $x=0$) is presented in Fig. ~\ref{fig3_fit}. 1D (red circles) and 2D (gray triangles) geometries are compared from $\tau=0$ to $\tau=45$.
An important difference can be seen between the two geometries: in the 2D situation the differential pressure becomes negative at $\tau=20$ as it goes to zero in the 1D case.
The observation of the negative pressure is the typical signature of a blast wind.
This measurement reveals the dramatic impact of the geometry on blast wind in a fluid of light and exemplifies the analogy with classical hydrodynamics.
To quantify this analogy, we use the Friedlander waveform model which is known to describe the dynamics of physical quantities in a free-field (i.e. in a open 3-dimensional space) blast wave \cite{Dewey2016}.
In this model the differential pressure follows an exponential decay of the form:
\begin{equation}
    \Delta \tilde{P}=P_s e^{-\tau/t^*}(1-\tau/t^*),
\label{friedland}
\end{equation}
where $P_s$ and $t^*$ are  two parameters which corresponds respectively to the peak differential pressure immediately behind the shock and to the time when the differential pressure becomes negative.
The period when the hydrostatic pressure is above the ambient value is known as the positive phase, and the period when the properties are below the ambient value is the negative phase. 
We use $P_s=1$ (since the differential pressure is normalized) and $t^*=20$ and plot the corresponding model with a black dashed line in Fig.~\ref{fig3_fit}.
An intriguing feature can also be seen in the 2D time evolution at $\tau=40$.
Close to the minimum of the negative phase, a second peak of differential pressure is observed (the single point at $\tau=40$ Fig.~\ref{fig3_fit} is the average of several realizations with errors bars indicating the standard deviation of the measurement) in our optical analogue which is reminiscent of the second shock observed in classical explosion.
In classical blast wave dynamics, this second shock is believed to be a consequence of the expansion and subsequent implosion of the detonation products and source materials.
Our results suggest that this second shock might be of more general nature than currently thought.

%Using Eq. (\ref{OverPressure}) the over-pressure due to the perturbation is calculated and averaged (over the neighboring slice of the transverse direction in the 1D case and radially in the 2D case) as shown on fig.~\ref{fig2_pressure}a,c. 

%Discussion: negative pressure clearly evidenced in the 2D case

\section*{Velocity}
\begin{figure}
\includegraphics[width=1\columnwidth]{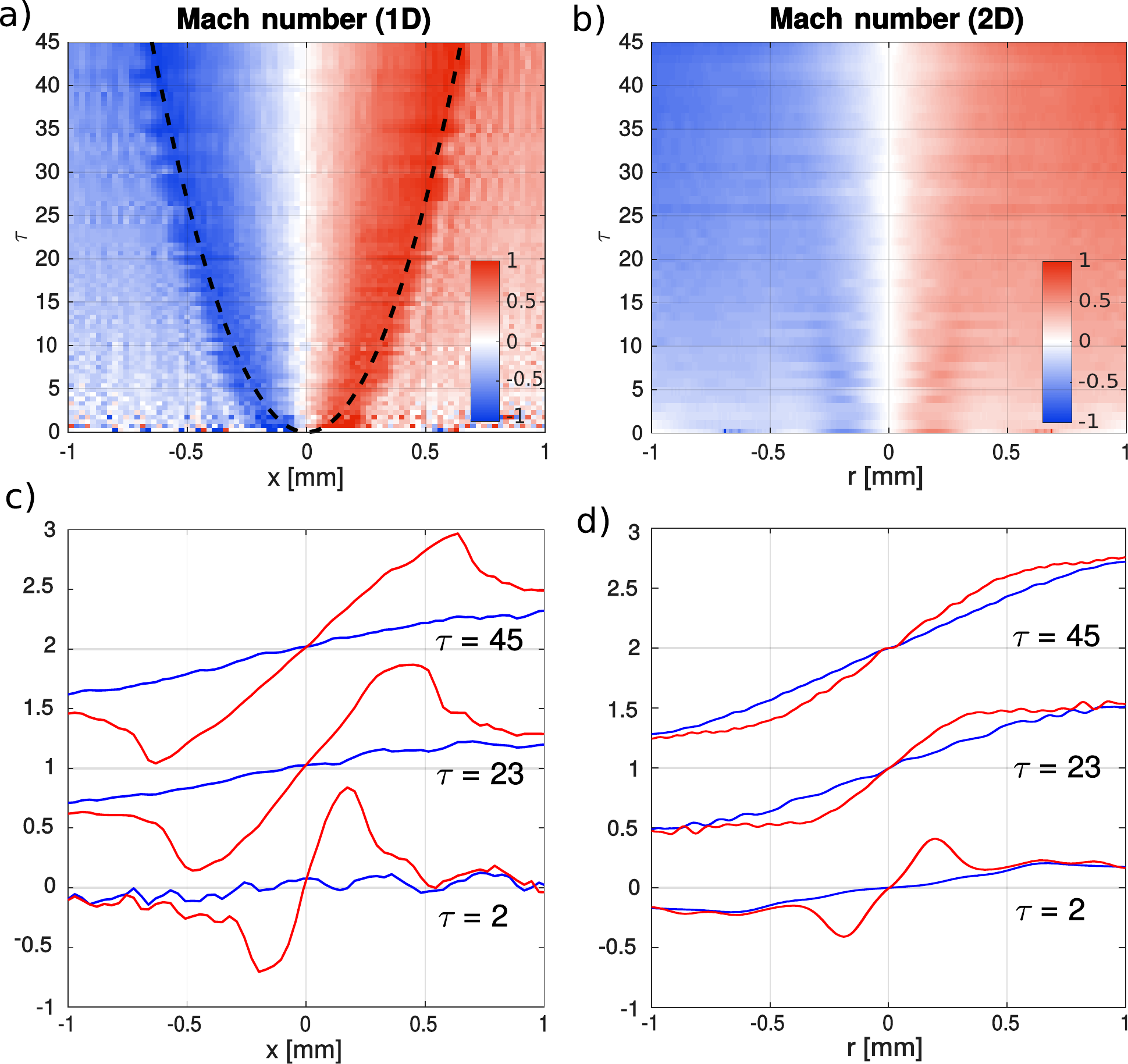}
\caption{Fluid velocities from the off-axis interferometry. 
a),b) Space-time evolution of the Mach number with respect to the background's local speed of sound, in the 1D and 2D geometry, respectively. The dotted black line in a) shows the calculated trajectory of expansion at the speed of sound (see main text).
c),d) show the background's $\tilde v_0$ (blue) and total $\tilde v$ (red) Mach number profiles, at different times, for the 1D (x coordinate) and 2D geometry (radial coordinate), respectively. 
Each following profile shifted vertically (spacing of 1) for  visibility. 
} 
\label{fig4_phase}
\end{figure}
%disucss figure 4, how to, and explain it is a useful tool for future expt .
For blast waves, there are no simple thermodynamic relationships between the physical properties of the fluid at a fixed point \cite{voronov1992nonlinear}. 
This means that the temporal evolution of the static pressure measured at a fixed point is not sufficient to calculate the temporal evolution of the velocity or the dynamic pressure from that single measurement.
To fully describe the physical properties of a fluid in a blast wave it is necessary to independently measure at least three of the physical properties, such as, the static pressure, the density and the fluid velocity or the dynamic pressure.
In the last section of this work, we report the measurement of last two physical properties, which are vector quantities.

The fluid velocity is calculated from its phase (see Eq. (\ref{FlVelo})) which is measured  using off-axis interferometric imaging.
The off-axis configuration consists in the tilted recombination of the signal beam with the reference beam on the camera plane.
This results in the set of linear fringes evolving along the relative tilt direction and locally deformed (stretched or compressed) according to the beams relative curvature.
Using a collimated Gaussian beam as the reference, the measured curvature is the one of the signal beam.
The acquired interferogramm carries the information on the beam phase via its amplitude modulated term. 
This term shows spatial periodicity and in the Fourier space it translates to two peaks shifted by a distance proportional to the off-axis tilt angle, symmetric with respect to the origin. 
By numerically calculating the spatial spectrum and filtering one of these peaks, the inverse Fourier transform gives the beam complex envelope with a spatial resolution bound by the fringe wavelength. 
The measured phase is unwrapped and the contribution due to the relative tilt is removed by subtracting the phase ramp. 
The resulting phase is averaged and numerically differentiated to get the velocity map.

Using this procedure, the off-axis interferograms of the background fluid and of the background fluid with the perturbation are analyzed to give access to $v_0(r,\tau)$ and $v(r,\tau)$, respectively. 
The difference of these quantities gives the perturbation velocity $v_1(r,\tau)$.
The non-zero velocity $v_0$ of the background fluid arises from its finite size causing its expansion due to a non-zero pressure gradient.
The knowledge of $v_0$ is essential to calculate the effective interaction $g$ and therefore the time $\tau$ and the sound velocity.
Indeed, $\phi_0=\tau\tilde\rho_0$ can be accessed by integrating $v_0$ over the transverse coordinate and using the fact that $\phi(\tilde r\rightarrow \infty,\tau)\rightarrow 0$. 
Knowing $\tau$, the sound velocity is $c_s(\mathbf{r}_\perp,\tau)=c\sqrt{\tau\tilde\rho_0(\mathbf{r}_\perp,\tau)/(k_0 L)}$.

The velocity maps normalized by the local sound velocity (in Mach units) are presented in figure \ref{fig4_phase} a) and b) for the 1D and 2D configurations, respectively. 
Since velocity is a vector quantity, negative values correspond to a propagation along $-x$ direction.
Figure \ref{fig4_phase} c) and d) show the corresponding profiles obtained for three specific times $\tau= 2;\ 23$ and $45$.
The maximal speed of sound at these times is 0.18, 0.62 and 0.86 percent of the speed of light in vacuum.
Positive outward velocity, as well as zero velocity at the center is observed at all times both in the 1D and 2D cases. 
Whereas it is intuitively expected in the 1D geometry with the differential pressure never dropping to negative values, it also holds in the 2D case in which a negative phase for the differential pressure exists. 
A possible explanation lies in the fact that when the negative phase is reached for the differential pressure, the perturbation has already expanded enough such that the net resulting force is smaller due to a larger distance. 
It is also worth noting that the velocity is at least 2 times larger in the 1D geometry than in 2D, as seen by comparison of the y-axis scales in Figure \ref{fig4_phase} c) and d).
Additionally, clear steepening of the velocity profiles is observed in the 1D case reaching  a Mach number of 1 at the steepest position.                 

\begin{figure}[h!]
\includegraphics[width=0.96\columnwidth]{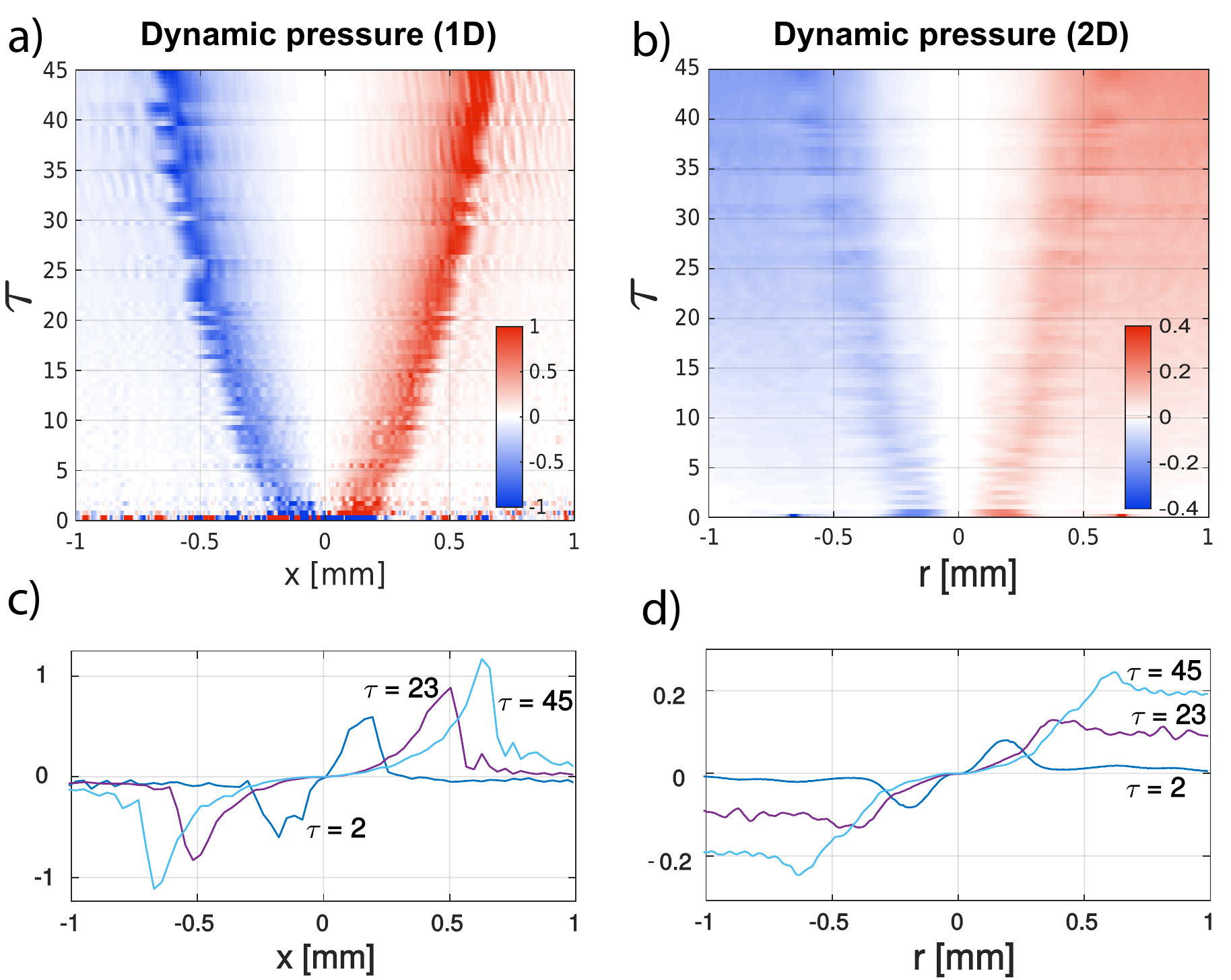}
\caption{
Dynamic pressure analysis. a) and b) show the spatio-temporal evolution maps of the dynamic pressure profiles, for the 1D (the x component) and 2D geometry (the radial component), respectively. Below, the c) and d) panels show various superimposed dynamic pressure profiles at different times, in 1D and 2D geometry, respectively.}
\label{fig5_pressureDyna}
\end{figure}

\section*{Dynamic pressure}
Alternatively, we can measure the dynamic pressure to compute a third thermodynamic quantity: the total pressure.
The dynamic pressure is also a vector quantity and can be obtained from a phase measurement similar to fluid velocity using Eq.~(\ref{PresDyna}).
The dynamic pressure maps are presented in Figs.~\ref{fig5_pressureDyna} a) and b). 
Once again Figs.~\ref{fig5_pressureDyna} c) and d) show dynamic pressure profiles for three selected times.
In 1D, the dynamic pressure forms a steep overpressure characteristic of the shock front which increases as function of time. 
In the 2D geometry, on the contrary the dynamic pressure reaches a plateau at the shock front without forming a steep overpressure peak.
This behavior is in agreement with the velocity distributions presented previously.

\section*{Conclusion}
Relying on detailed measurements of all thermodynamic quantities in a fluid of light blast wave, we have demonstrated for the first time the occurence of a blast wave in a fluid of light.
We compare 1D and 2D geometry and report the observation of a negative phase during the blast only for the 2-dimensional case.
The differential pressure in the 2D geometry is compared to the classical hydrodynamics of Friedlander blast-wave and we see a very good agreement with this model.
%We also observe a second shock reminiscent of TNT explosive blast wave.
Velocity maps and dynamic pressure are finally presented to complete the study.
Our work opens the way to precise engineering of a fluid of light density and velocity distribution which will prove to be a valuable tool to design new experiments studying superfluid turbulence \cite{rodrigues2020turbulence}
or analogue gravity where an excitation of a fluid of light changes from a subsonic to a supersonic region.\\
~

\acknowledgments

The authors thank Ferdinand Claude and Samuel Deléglise for useful discussions for setting up PyRPL.
This work is supported by the PhoQus project.

%\nocite{*}

\bibliography{biblio}% Produces the bibliography via BibTeX.

%%%%%%%%%% Merge with supplemental materials %%%%%%%%%%
\clearpage
\widetext
\begin{center}
\textbf{\large Supplemental Materials: Blast waves in a paraxial fluid of light}
\end{center}
%%%%%%%%%% Merge with supplemental materials %%%%%%%%%%
%%%%%%%%%% Prefix a "S" to all equations, figures, tables and reset the counter %%%%%%%%%%
\setcounter{equation}{0}
\setcounter{figure}{0}
\setcounter{table}{0}
\setcounter{page}{1}
\makeatletter
\renewcommand{\theequation}{S\arabic{equation}}
\renewcommand{\thefigure}{S\arabic{figure}}
\renewcommand{\bibnumfmt}[1]{[S#1]}
%\renewcommand{\citenumfont}[1]{S#1}
%%%%%%%%%% Prefix a "S" to all equations, figures, tables and reset the counter %%%%%%%%%%

\begin{figure}[h]
\includegraphics[width=0.9\columnwidth]{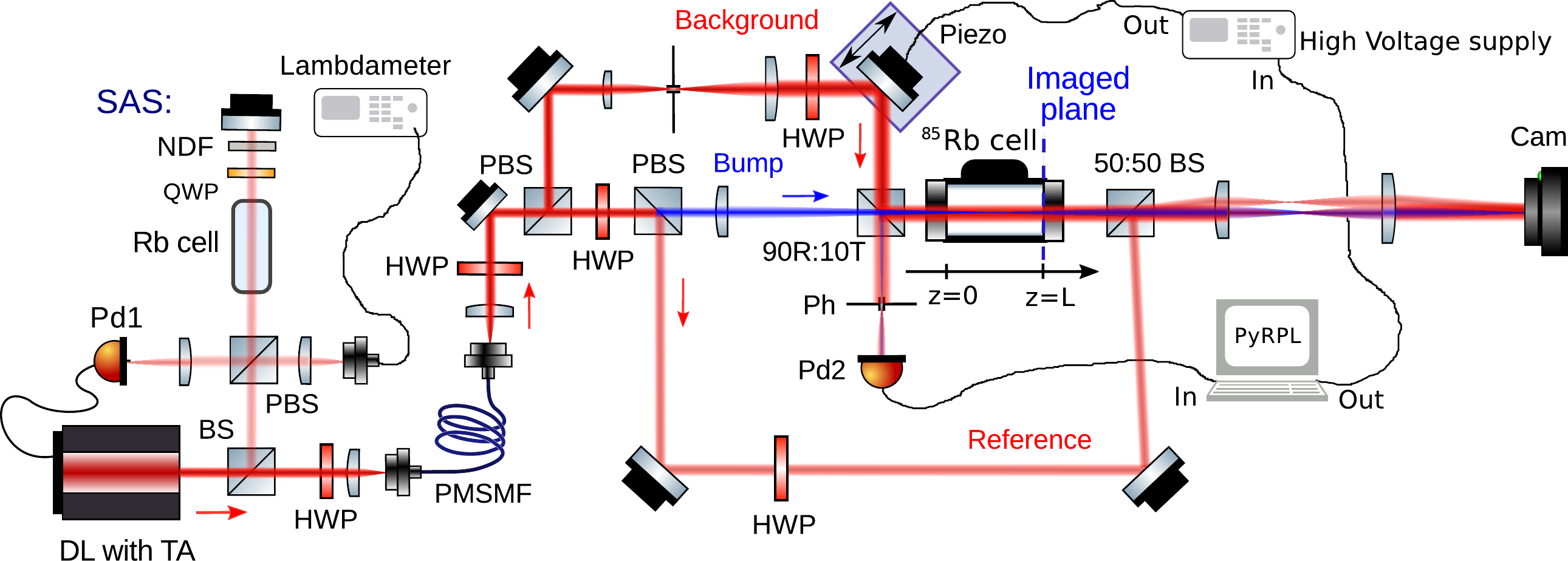}
\caption{
Schematic visualization of the experimental setup. Diode laser frequency calibration was performed with Saturable Absorption Spectroscopy (SAS), and during the experiment the frequency measurement was performed with a MogWave Lambdameter. 
The laser was mode cleaned with a polarization maintaining single mode siber (PMSMF), before being split into the Background, Bump and the Reference. 
The Background-Bump interference arm complementary to the Rb vapor cell was cropped with a 200~$\mu m$ diameter pinhole (Ph) to measure the the power of the overlap area on a photodiode. 
This signal was minimized by controlling the relative phase via piezoelectric motion of a mirror mount to have permanently constructive interference on the vapor cell arm. 
The error signal was generated from the photodiode signal with the PyRPL lockbox software.}
\label{Setup}
\end{figure}

\textbf{Experimental details} 
The scheme of the experimental setup is shown on Figure \ref{Setup}. 
Toptica DLCpro 780 with TA was used for all measurements. The laser frequency was tuned around 780~nm and measured with a MogWave Multimeter LambdaMeter and calibrated with Saturable absorption spectroscopy (SAS). 
The laser beam was mode cleaned with a single mode fiber and then split into the Background, Bump and the Reference arms. 
The respective intensity ratio was fixed by the angles of the Half-Wave-Plates (HWP), placed before the Polarizing Beam Splitters (PBS), in agreement with experimental requirements: $\tilde\rho_1(\textbf{r}=0,\tau=0)\approx3$ and minimal sufficient power into the reference beam to have noticeable fringe contrast.
Since the Background and the Reference have the same polarization during recombination, their interference needs to be constructive at the cell input in order to create the desired input state for the fluid's density.
The beamsplitter's unused arm's power at the Background-Bump overlap area should then be minimal.
This signal was measured with a 200~$\mu m$ diameter pinhole centered at the overlap area and a photodiode.
The relative phase needs to be locked in order to minimize permanently this signal and make it insensitive to perturbations such as air currents. 
Therefore the photodiode signal was transformed into an error signal of a piezoelectric mirror mount controlling the relative phase. 
The error signal generation from the photodiode signal was realized with the PyRPL software running on a Red Pitaya FPGA \cite{PyRPLref}.
The modulation frequency was around 2-3 kHz.

~\\

\textbf{Vapor Temperature} 
\begin{figure}
\includegraphics[width=0.8\columnwidth]{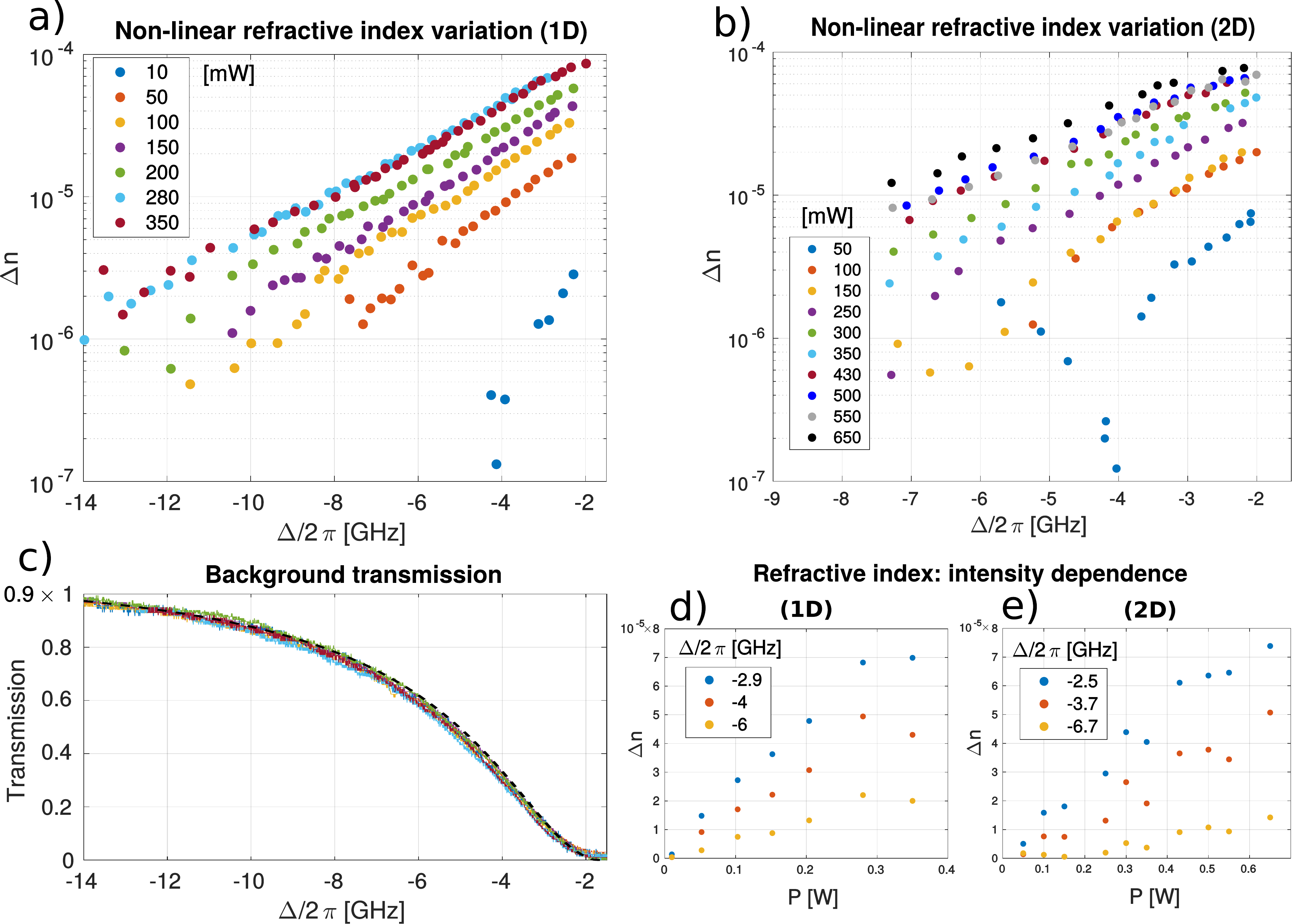}
\caption{
Vapor's transmission and its intensity dependent refractive index measurement.
a) and b) show the maximal refractive index variation  calculated from the off-axis interferograms of the backgroung beam with a reference, in 1D and 2D geometry, respectively. 
c) Background beam's transmission spectrum with respect to $^{85}$Rb cooling transition measured at different input powers. 
Dashed line is the theory of a linear multilevel vapor at temperature 150~°C and 0.5~\% the isotopic fraction of $^{87}$Rb inside the cell. 
Checking the "Kerr" approximation:
d) and e) show the variation of the refractive index with laser power at fixed laser detuning in both geometries. }
\label{fig_Dn}
\end{figure}

One of the useful knobs to control the light-matter interaction in hot vapor cells is the atomic density. 
The latter is directly linked to the vapor pressure via the ideal gas law (neglecting the atom-atom interactions). 
It equals the Rb vapor's saturation pressure at thermal liquid-gas equilibrium and can be increased by several orders of magnitude when heating the cell from 50°~C to 150°~C. 
Keeping the vapor temperature constant during the experiment is therefore necessary to control the atomic susceptibility.
In our experiment, several electric resistors were wound around the cell and connected in parallel to a DC power supply to heat up the cell. 
The vapor temperature was accessed by measuring the transmission spectrum around the Rb D2 line in the weak beam limit. 
The frequency calibration was performed via Saturable absorption spectroscopy, as shown on Figure \ref{Setup}. 
The experimental spectrum was fitted with the linear susceptibility model developped in \cite{siddons2008} taking into account all hyperfine transitions of both isotopes and the collisional self-broadening due to resonant dipole-dipole interactions \cite{Weller_2011}, with the atomic density and the number fraction of $^{87}$Rb isotope as free parameters. 
The temperature was measured before and after each experiment to prevent any temperature drift. 
\\ \\
\textbf{Non-linear refractive index variation measurement}
The intensity dependent refractive index of our hot atomic vapor is the key parameter governing the fluid's dynamics as it is linked to the effective evolution time: $\tau=\Delta n k_0L$ and its speed of sound: $c_s=c\sqrt{\Delta n}$.
In this work it was measured using the off-axis interferometry which gives access to the transverse phase variations at the cell exit plane. The transverse phase profile of the Background beam is assumed to depend as follows on the beam's intensity I(\textbf{r}):
\begin{equation}
    \phi_{th}(\textbf{r},L)=k_0L\frac{n_2I(\textbf{r})}{1+I(\textbf{r})/I_s}+\phi_0
\end{equation}
Where $n_2$ is the Kerr index, $I_s$ the saturation intensity of the Kerr effect and $\phi_0$ a constant phase. The gradient of the phase, giving access to the fluid velocity, is numerically calculated and fitted with $\nabla\phi_{th}$ with $n_2$ and $I_s$ as free parameters. Figure \ref{fig_Dn} a) and b) show measured maximal variation of refractive index for different experimental configurations of the laser detuning $\Delta$ and power $P$. Each point corresponds to a processed image. c) Shows the transmission spectra through the cell for different input powers. 
No saturation of the absortpion can be evidenced. 
The black dashed line is the theoretical calculation of the linear susceptibility used for the measurement of the vapor's temperature.
Finally, d) and e) show the variation of the refractive index with intensity. 
The graphs show that the results of this work are obtained below the regime of the saturation of the Kerr effect.
\\
\textbf{Background beam's expansion}
\begin{figure}
\includegraphics[width=0.55\columnwidth]{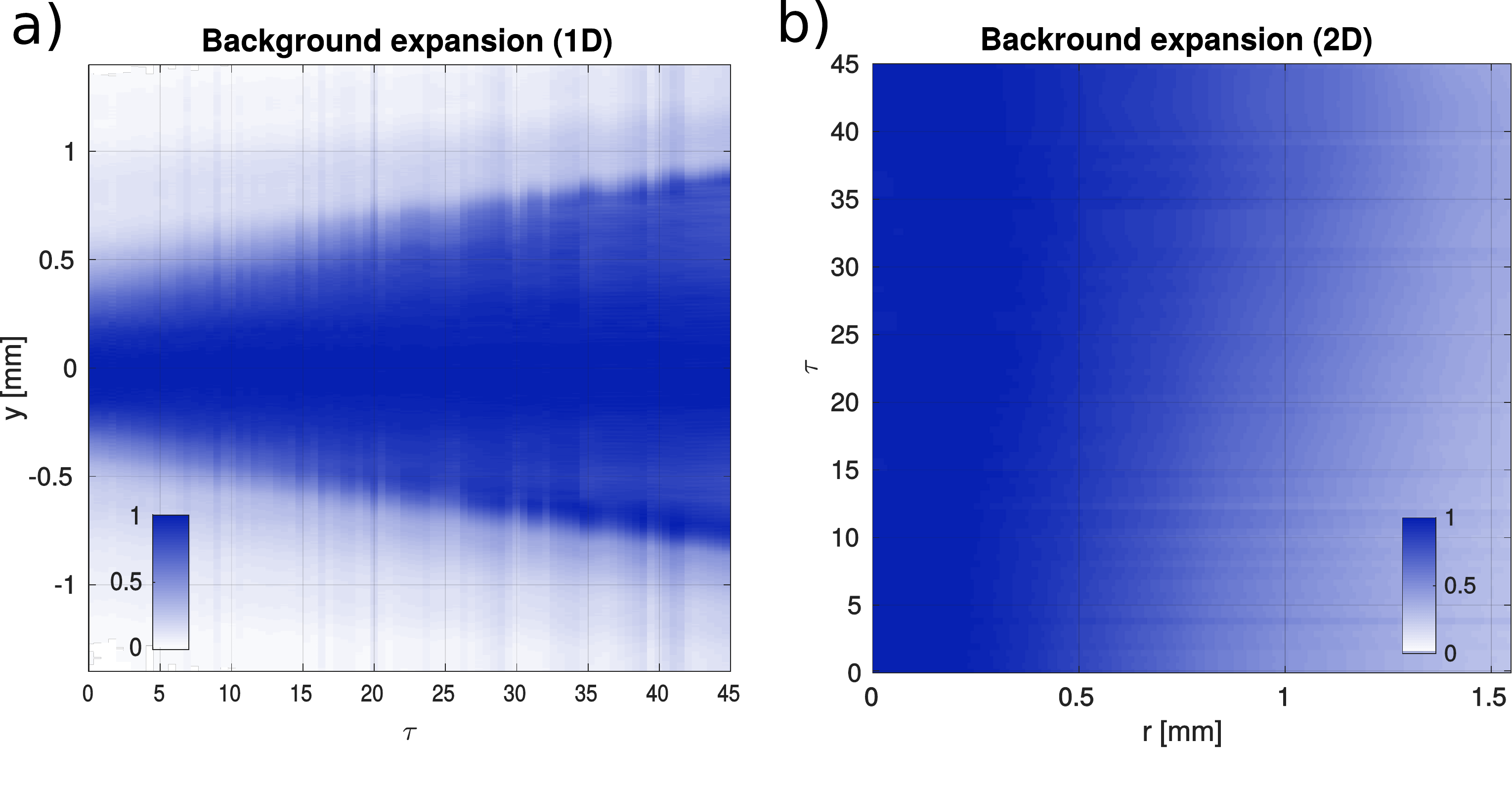}
\caption{
Background fluid's expansion. a) shows the Background's expansion in the transverse y direction in the 1D geometry and b) shows the Background's radial expansion in the 2D geometry.}
\label{fig_expansion}
\end{figure}

In the theoretical discussion developed in the main text and for the $\Delta n$ measurement it is assumed that the background fluid beam's density is invariant with time. The experimental data to verify this hypothesis are shown in Figure \ref{fig_expansion}. No expansion in the x direction of the 1D case was observed. The expansion is most pronounced in the transverse y direction of the 1D case. In the 2D case the background's insignificant expansion is observed.
\\
\textbf{Relevance of the Quantum Pressure}
\begin{figure}[]
\includegraphics[width=0.6\columnwidth]{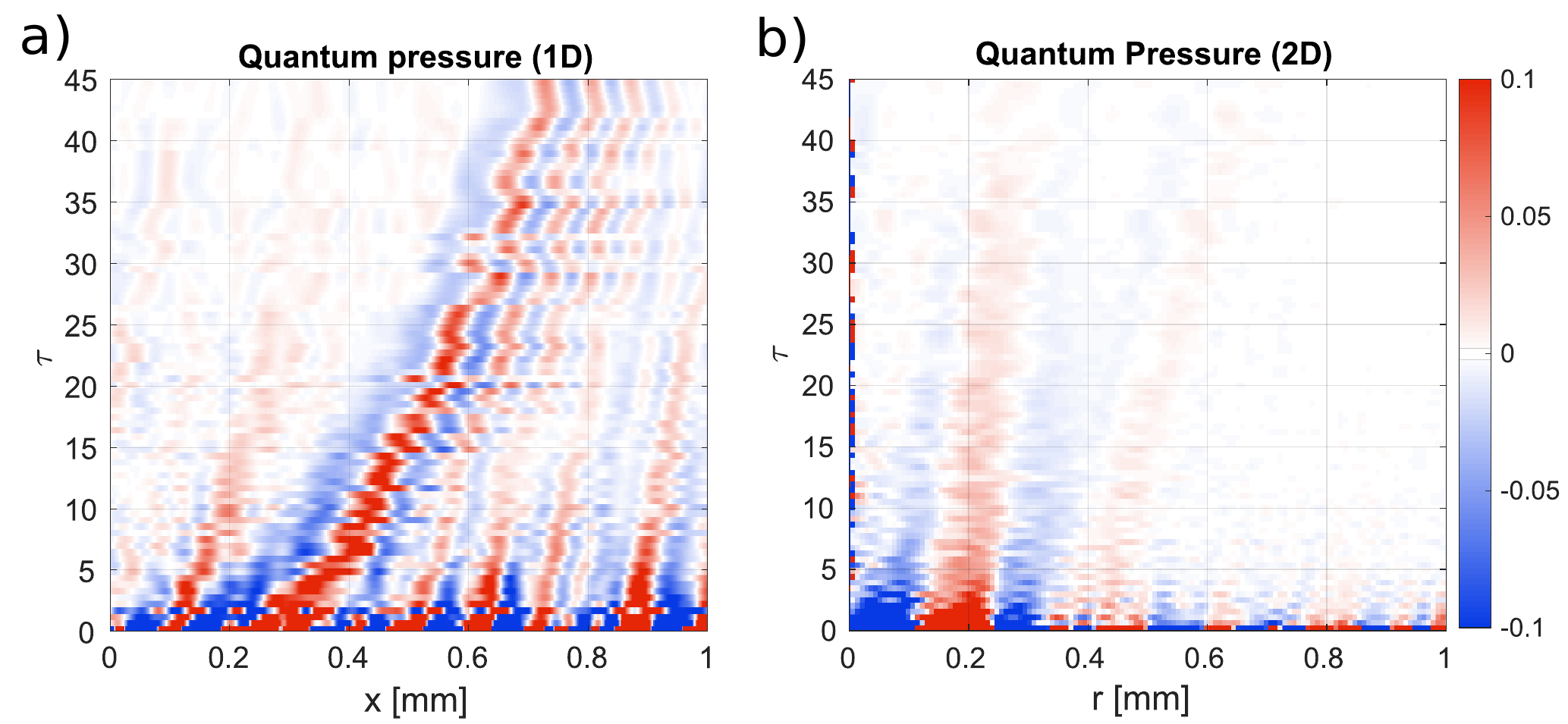}
\caption{
Quantum pressure calculated from experimental density profiles. a) in the 1D geometry and b) in the 2D geometry. Same colormap is used for both graphs.}
\label{fig_QP}
\end{figure}

As mentioned in the main text, the Quantum pressure was neglected in the theoretical description of the experimental data as we are interested in the fluid's behavior in the long wavelength limit. 
This term is known to have a dispersive contribution to the shockwave profile which, upon steepening, becomes composed of an increased amount of various momentum components moving at different velocities.
To evaluate the relevance of the Quantum Pressure in this work we calculated it from the experimental density profiles at different evolution times for both 1D and 2D geometry as:
\begin{equation}
    \tilde P_q = \frac{1}{2\sqrt{\tilde\rho}}
    \tilde\nabla^2_{\perp}\sqrt{\tilde\rho}
\end{equation}
Depending on the dimensionality the Laplacian was calculated as: $\tilde\nabla^2_{\perp}=\xi^2 \partial^2/\partial x^2$ in 1D or as: $\tilde\nabla^2_{\perp}=\xi^2[ \partial^2/\partial r^2+(1/r)\times\partial/\partial r]$ in 2D using the radial symmetry.
With this formulation the value of the dimensionless Quantum Pressure directly compares with the dimensionless density (stemming from interactions) in the right hand side of the Euler-like Madelung equation.
The result is shown on Figure~\ref{fig_QP} a) for the 1D and b) for the 2D case. 
The Quantum Pressure seems to be most pronounced at the vicinity of the Shock front in 1D.
In 2D it seems to decay with time. In both cases it does not exceed 0.1 for times $\tau>5$.
This validates the theoretical approach chosen in work and consisting in neglecting the Quantum Pressure.
For lower times the calculation seems inaccurate. This may be due to a large uncertainty on the healing length $\xi$ in this regime.

\end{document}